\documentclass{article}

\usepackage{graphicx}  
\title{Heavy quarks, from discovery to precision\protect\footnote{Talk given at
``Rencontres de Physique de la Vall\'ee d'Aoste'', La Thuile, Italy, March 2011, in
honour of Mario Greco's 70$^\mathrm{th}$ birthday.}}
\author{Matteo Cacciari\\
\small
LPTHE, UPMC and CNRS, Paris, France\\
\small
Universit\'e Paris Diderot, France
}
\date{}

\begin{document}

\maketitle

\begin{abstract} 
The discoveries of the heavy quarks are briefly reviewed, with a focus
on the role played by Mario Greco in the interpretation of the experimental
observations, and on his contributions to heavy quark precision
phenomenology. 
\end{abstract}

\section{Mario's charm}

In November 1974 two experimental groups simultaneously announced the discovery of
a new resonance. The collaboration led by Sam Ting \cite{Aubert:1974js} at the
Brookhaven National Laboratory and the one led by Burton Richter
\cite{Augustin:1974xw} at the Stanford Linear Accelerator Laboratory agreed on all
the key characteristics of the new particle, but its name. Since the latter is not
consequential, we shall rather focus here on its mass, at 3~GeV significantly
larger than previously observed hadronic resonances and -- more importantly -- its
total width, estimated at less than 1.3~MeV in \cite{Augustin:1974xw}, a
surprisingly small value for a hadronic resonance. Appelquist and Politzer
\cite{Appelquist:1974zd} and De Rujula and Glashow \cite{De Rujula:1974nx} are
credited with the first interpretation of the new particle (eventually called
$J\!/\!\psi$) as a bound state of the previously unobserved charm quark and its
antiquark. The relatively large mass of the new quark ($\sim$~1.5~GeV), together
with the asymptotic freedom property of QCD, could elegantly explain the very
small observed width.

\begin{figure}[t]
\begin{center}
\includegraphics[width=\textwidth]{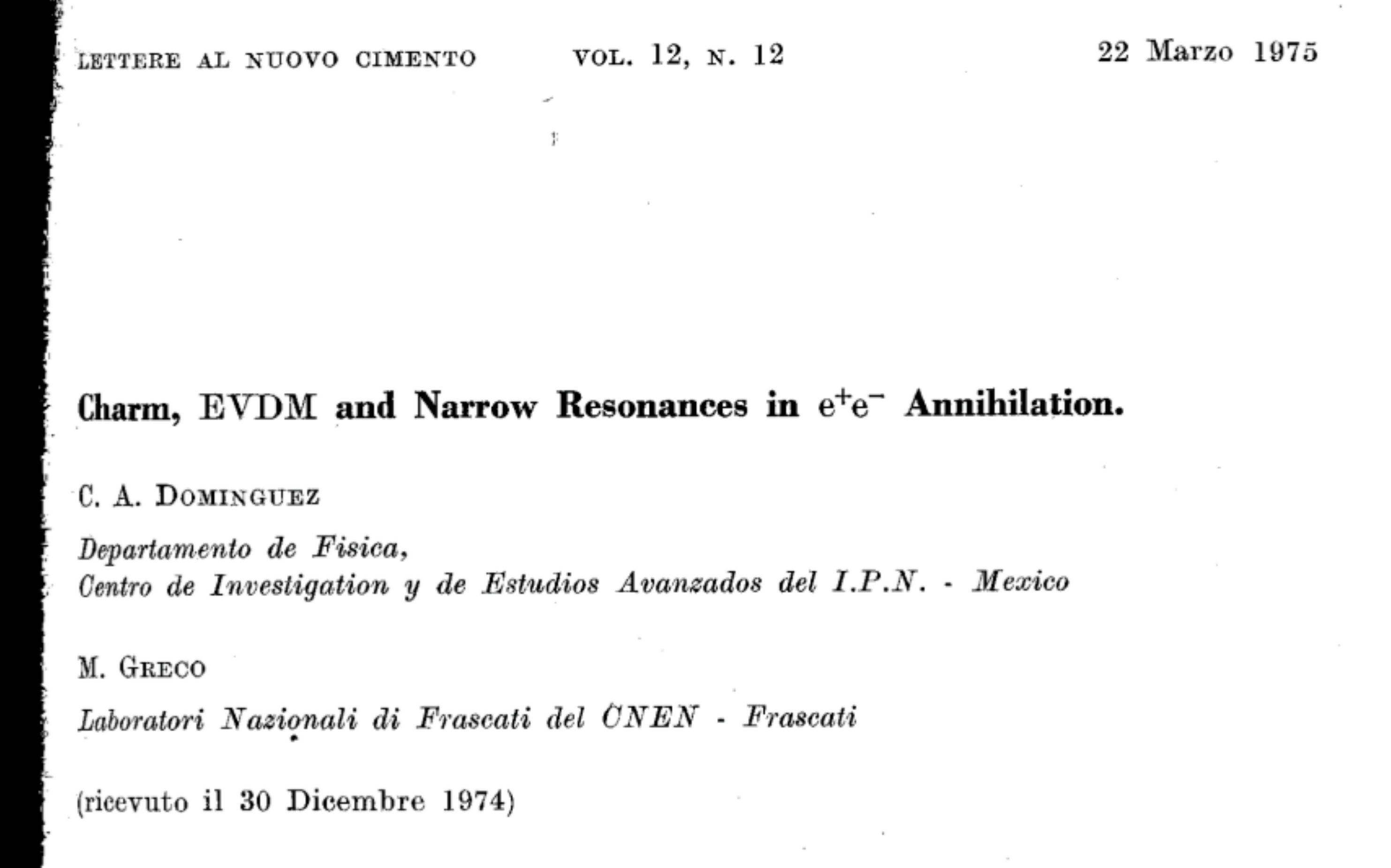}     
\caption{\label{fig:greco-charm} The front page of the Dominguez-Greco
\protect\cite{Dominguez:1974be} paper interpreting the observation of the $J\!/\!\psi$ as
a charm-anticharm vector bound state. The typo in the title, EV{\bf DM} rather than
EV{\bf MD} (a clear indication of how hectic those times must have been), bears fortunately no relation with the accuracy of the paper.}
\end{center}
\end{figure}

Mario Greco was 33 years old and en route to SLAC for a seminar
when the news of the discovery broke. Once at destination he was able to
gather the available details, notably the mass of the resonance, and forward
them to Frascati, where the observation could immediately
be confirmed by the ADONE $e^+e^-$ collider~\cite{Bacci:1974za}.
Mario then flew to Mexico City for a planned visit, and once there 
he learnt about the discovery of the $\psi'$ through the local press. In collaboration
with C.A.~Dominguez  he quickly published a paper \cite{Dominguez:1974be}. Working 
within the Extended Vector Meson
Dominance (EVMD) approach \cite{Bramon:1972vv}, and using the scarce experimental
data available about the new $\psi_n$ resonances, they were able to derive their
 total
contribution to hadron production in $e^+e^-$ collisions. They wrote:
\begin{eqnarray}
R &=& \frac{\sigma(e^+e^- \to \gamma\to\mathrm{hadrons}) + 
\sigma(e^+e^- \to \psi_n\to\mathrm{hadrons})}{
\sigma(e^+e^- \to \gamma\to\mu^+\mu^-) + 
\sigma(e^+e^- \to \psi_n\to\mu^+\mu^-)} \nonumber\\
&=& R_\mathrm{normal} + R_\mathrm{charm}
\simeq 2.5 + 1.2 = 3.7
\end{eqnarray}
The resulting increment for the $R$ ratio was in fair agreement
with experimental data, and allowed them to interpret the newly observed
resonances:``{\it ...one is naturally led to think of the new narrow
resonances as charm-anticharm vector mesons.}''

\section{Mario's beauty}

A few years later it was the turn of another quark to make its appearance in the
form of a new resonance. In 1977 the collaboration led by Leon Lederman observed a peak
around 9.5 GeV in the structure of the dimuon spectrum in 400 GeV proton-nucleus
collisions at the Fermilab \cite{Herb:1977ek}. This was quickly interpreted as a
bottom (or beauty)-antibottom bound state. Shortly thereafter, Mario Greco applied
again \cite{Greco:1977tg} duality
ideas \cite{Bramon:1972vv,Greco:1973wm,Boehm:1973pi,Sakurai:1973rh,Gounaris:1974tm,Etim:1974cp}
to this discovery. These ideas led to simple relations for the electronic widths of
vector mesons,
\begin{equation}
\Gamma^{e\bar e}_\rho : 
\Gamma^{e\bar e}_\omega : 
\Gamma^{e\bar e}_\varphi : 
\Gamma^{e\bar e}_\psi : 
\Gamma^{e\bar e}_\Upsilon = 9 : 1 : 2 : 8 : 2(8)
\end{equation}
where the last term in the equation above is related to the electric charge of the
bottom quark having the value $-1/3$(2/3). Choosing the value $-1/3$ leads to the
prediction $\Gamma^{e\bar e}_\Upsilon \simeq 1.2$~keV. This, in turn, allows one
to estimate the production cross section of the $\Upsilon$, for which Mario
obtained a value in good agreement (within a factor of two) with the experimental
measurement. He could therefore conclude that the charge $-1/3$ for the bottom quark
was favoured by the available data: ``{\it Our results suggest that the charge of
the new constituent quark is likely $-1/3$}''.

A by-product of this analysis were the predictions for the values for the leptonic widths of the
$\Upsilon$ and the higher resonances, at the time unknown. Table \ref{table1} compares the predictions in 
\cite{Greco:1977tg} with the modern measured values.
\begin{table}
\begin{center}
\begin{tabular}{|c|c|c|c|}
$\Gamma^{e\bar e}$ (keV) & $\Upsilon$  & $\Upsilon'$  & $\Upsilon''$ \\
\hline
Mario Greco \cite{Greco:1977tg}      & 1.2 & 0.65 & 0.55 \\
PDG \cite{Nakamura:2010zzi} & 1.34& 0.61 & 0.44 \\
\end{tabular}
\caption{\label{table1} The predictions of ref.~\cite{Greco:1977tg} for the
electronic widths of bottom-antibottom bound states, compared to modern experimental
results.}
\end{center}
\end{table}
Obviously, not a bad job.

\section{Top discovery}

After these two discoveries almost twenty years elapsed before the sixth quark was
finally observed. The CDF collaboration at the Fermilab Tevatron collider published
at first initial evidence \cite{Abe:1994st} for the top quark in 1994, and followed
up in 1995 with the definitive observation \cite{Abe:1995hr}. This last paper was
also  presented \cite{Gerdes:1995qj} in the 1995 edition of the La Thuile
conference,  one of the very first public announcements of the definitive
discovery of the top quark.

\begin{figure}[t]
\begin{center}
\includegraphics[width=0.8\textwidth]{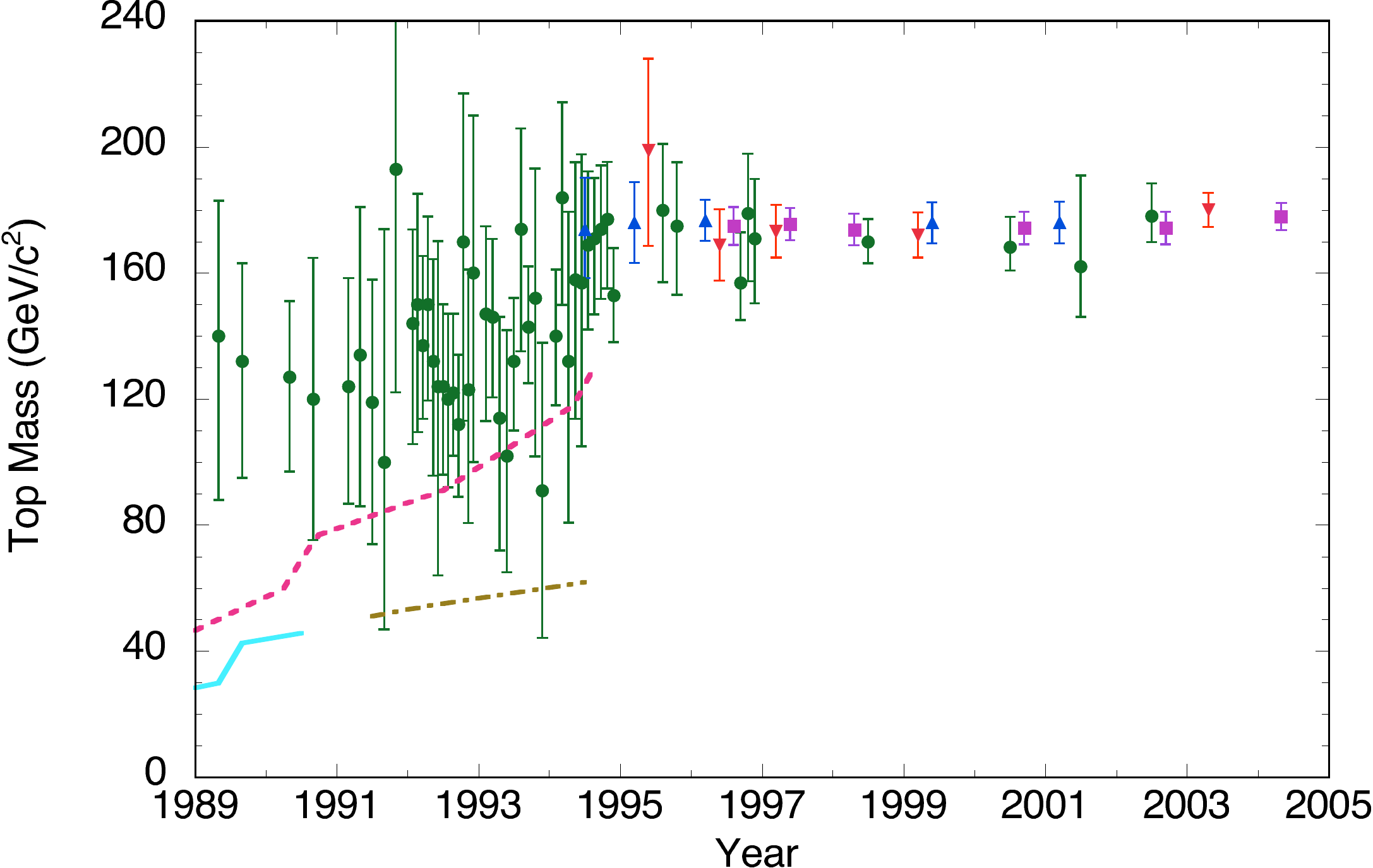}     
\caption{\label{fig:topmass} The evolution in time of the top mass value extracted from
electroweak precision fits at LEP (green circles), together with the actual
measurements at the Tevatron (red and blue triangles, magenta squares). 
Taken from ref.~\cite{Quigg:2004is}, page 24, fig. 16.}
\end{center}
\end{figure}

The very large mass, of the order of 175 GeV, at which the top quark was finally
observed would have been perhaps surprising only a few years earlier when, without
any other experimental guidance, one could have expected a top quark only
marginally heavier than the heavy quarks already discovered. However, by the time
of the CDF discovery, a lot more information was available through the precision
fits of the Standard Model parameters performed at LEP. In particular, it had
become clear (see e.g. fig.~\ref{fig:topmass}, taken from \cite{Quigg:2004is}) that
the top quark was going to be very heavy, with a mass of the order of 150 GeV, and
a residual uncertainty that, in 1994, was probably of the order of $\pm 20$-$30$
GeV. This indirect evidence for the value of the top mass was one of the main
contributions of LEP to the experimental landscape, and it was possible because of
a huge amount of theoretical and phenomenological work directed at improving the
predictions. As an example of Mario Greco's contribution to this collective effort
I'd like to mention two of his many papers on radiative corrections for LEP
physics, refs.~\cite{Greco:1980mh} and  \cite{Consoli:1982ib}, which extensively reviewed 
and systematized electromagnetic corrections to Bhabha scattering at the $Z^0$ pole.

\section{`Precision' physics in heavy quarks and quarkonium}

After the time of discoveries comes of course that of more accurate measurements
and, out of necessity, more refined theoretical predictions, usually in the form of
next-to-leading order (NLO) and resummed calculations. I wish to mention in
particular two contributions of Mario Greco to this endeavour.

On of them is the first complete and systematic NLO calculation of heavy quarkonium
total cross sections in hadronic collisions \cite{Petrelli:1997ge} within the then
recently developed Non-Relativistic QCD (NRQCD) formalism~\cite{Bodwin:1994jh}. This
work capped a series of papers on heavy quarkonium that Mario and I wrote together,
the first of them, on the role of resummed fragmentation contributions in the
production of $J\!/\!\psi$ at the Tevatron \cite{Cacciari:1994dr}, as part of my
doctoral thesis. Twenty years after its discovery, the $J\!/\!\psi$ was still
providing theorists with a lot of work, the focus having shifted to a
detailed understanding of its production mechanism and to accurate evaluations of its
cross sections, a quest that still goes on today.

\begin{figure}[t]
\begin{center}
\includegraphics[width=0.52\textwidth,angle=-90]{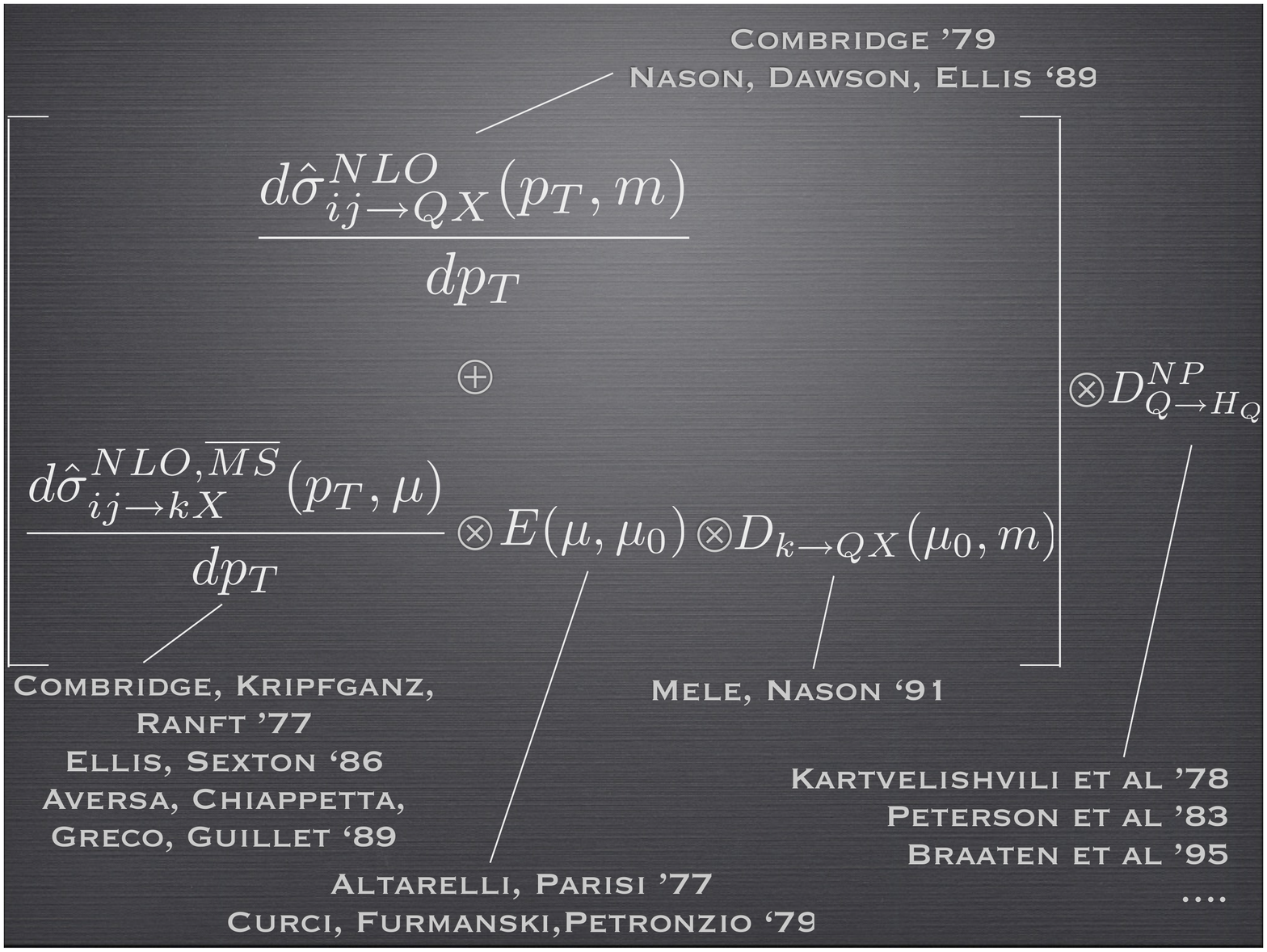}     
\caption{\label{fig:fonll} Schematic view of `previous art' used in the 
FONLL formalism, showing the
authors of the main ingredients that enter the calculation.}
\end{center}
\end{figure}

A second contribution of Mario to precision phenomenology is the large transverse
momentum resummation of heavy quark production in hadronic collisions
\cite{Cacciari:1993mq}, a paper that we wrote together in 1993 and my first foray
into QCD. At the time I was a graduate student in Pavia. Mario, who eventually
spent three years there, had just moved from a position with the INFN (the Italian
Institute for Nuclear Physics) to a professorship in the University. He suggested
that I look into combining the results of an article he had written a few years
earlier with Aversa, Chiappetta and Guillet, the full set of higher order QCD
corrections to parton-parton scattering processes \cite{Aversa:1988vb}, with those
from a paper from Mele and Nason \cite{Mele:1990cw}, which calculated the boundary
conditions of the fragmentation functions of massless partons into a massive
quark. Together with the evolution kernels from
Altarelli-Parisi~\cite{Altarelli:1977zs} and
Curci-Furmanski-Petronzio~\cite{Curci:1980uw}, these ingredients were what was 
needed to perform the resummation to next-to-leading
logarithmic level of the cross section for heavy quark production  at large transverse momentum.  The
availability of all the building blocks did not make the job look less daunting. Mario
put me in touch with Jean-Philippe Guillet and with Paolo Nason (and later Michel
Fontannaz), who kindly
provided us with codes they had written for other projects but which contained the
necessary ingredients. Then, patiently and with a keen understanding of what the
correct outcome had to look like, he helped me make sense of a  few thousand lines of
CAPITALISED Fortran 77 code and eventually obtain physically meaningful  results.

This work, also a part of my PhD
thesis,  has successively evolved into the so called FONLL calculation
\cite{Cacciari:1998it} of heavy quark production, a formalism where the fixed order
calculation at NLO~\cite{Nason:1989zy} is matched with the resummed one from
\cite{Cacciari:1993mq} and, at the same time, non-perturbative information
extracted from LEP data is employed in predictions of heavy hadrons spectra in
hadronic collisions. A schematic view of the FONLL calculation, in the form
\begin{equation}
d\sigma^{FONLL}_{H_Q} = [d\sigma^{NLO}_Q \oplus d\sigma^{res}_Q] \otimes D^{non-pert}_{Q\to
H_Q} \,
\end{equation}
where $\oplus$ denotes a `matched' sum and $\otimes$ a convolution, is given in 
figure~\ref{fig:fonll}. It shows how FONLL draws from a large amount of previous work
in QCD, 
achieving a remarkable synthesis. Eventually, this synthesis also proved to be
quite effective, as it was shown capable of describing well heavy quark
production in a number of different experiments, from $ep$ collisions at HERA, to
$pp$ and $p\bar p$ at RHIC and the Tevatron and, more recently and almost 20 years
after it was first introduced, $pp$ collisions at the LHC.

\section{Conclusions}

The history of heavy quarks is now almost forty years long, and Mario Greco's career
spanned all of it. His work has given many contributions to our present understanding,
and in these proceedings I could only describe briefly some of it.

The very much abridged story of these forty years started here with the discovery
of the fourth quark, charm. It may be easy, today and from the heights of our six
known quarks, the heaviest of them with potential links to new physics beyond the
electroweak scale, to take this fourth, barely `heavy' quark almost for granted.
This would however mean doing injustice to the revolutionary proposal of Glashow,
Iliopoulos and Maiani~\cite{Glashow:1970gm} which in 1970, introducing the charm
quark, presciently captured the lepton-hadron symmetry which is now a cornerstone
of the Standard Model. Indeed, its importance did not quite go unnoticed at the
time, and Collins, Wilczek and Zee~\cite{Collins:1978wz} could for instance write,
in 1978 and before the Nobel prize effectively sealed the paternity of the Standard
Model, ``{\it... we specialize to the standard sequential
Weinberg-Salam-Glashow-Iliopoulos-Maiani model of weak interactions...}''.

\section*{Acknowledgments}

I wish to thank the Organisers of La Thuile conference for inviting
me to give this talk. More importantly, I wish to thank Mario for the physics he
has taught me
and for the attitude towards physics that I have tried to learn from him. It has
been a privilege to be his student and collaborator, and a real pleasure to work with
him.

\end{document}